# Observation of quantum-Hall effect in gated epitaxial graphene grown on SiC (0001)

T. Shen,[1,2] J. J. Gu,[1] M. Xu,[1] Y.Q. Wu,[1] M.L. Bolen,[1] M.A. Capano,[1] L.W. Engel,[3] and P.D. Ye [1, a)]

[1)] School of Electrical and Computer Engineering and Birck Nanotechnology Center, Purdue University, West Lafayette, IN 47907
[2)] Department of Physics, Purdue University, West Lafayette, IN 47907
[3)] National High Magnetic Field Laboratory, Tallahassee, FL 32310

(October 28, 2009)

Epitaxial graphene films examined were formed on the Si-face of semi-insulating 4H-SiC substrates by a high temperature sublimation process. A high-*k* gate stack on the epitaxial graphene was realized by inserting a fully oxidized nanometer thin aluminum film as a seeding layer, followed by an atomic-layer deposition process. The electrical properties of epitaxial graphene films are retained after gate stack formation without significant degradation. At low temperatures, the quantum-Hall effect in Hall resistance is observed along with pronounced Shubnikov-de Haas oscillations in diagonal magneto-resistance of gated epitaxial graphene on SiC (0001).

Graphene, a monolayer of carbon atoms tightly packed into a two-dimensional (2D) hexagonal lattice, has recently been shown to be thermodynamically stable and exhibits astonishing transport properties, such as an electron mobility of ~15,000 $cm^2/Vs$ and an electron velocity of ~$10^8$ cm/s at room temperature.[1] High-quality monolayer graphene has been obtained in small (tens of microns) areas by exfoliation of highly-ordered pyrolytic graphite (HOPG) and transferred onto $SiO_2$ substrates for further device fabrication [2,3]. However, this exfoliation process cannot form the basis for a large-scale manufacturing process. Recent reports of large-area epitaxial graphene by thermal decomposition of SiC wafers have provided the missing pathway to a viable electronics technology [4-9]. An interesting question that remains to be addressed is whether the electrical properties of epitaxial graphene on SiC are essentially same as those in exfoliated graphene films.[10-11] For example, the well-known quantum Hall-effect (QHE), a distinguishing feature of a 2D electronic material system, has not been observed up to now in epitaxial graphene.[4]

The advantage of epitaxial graphene for nanoelectronic applications resides in its planar 2D structure that enables conventional top-down lithography and processing techniques. Except for opening the bandgap by forming graphene nano-ribbons, an additional challenge for graphene based electronics is the formation of high-quality, ultrathin dielectrics with low interface trap density. A perfect graphene surface is chemically inert, which does not lend itself to conventional atomic-layer deposited (ALD) high-k dielectrics.[12-14] In this Letter, we report on ALD high-k gate stack integration on epitaxial graphene films by inserting a fully oxidized aluminum film as a seeding layer.[15] The gate stack formation does not degrade the electrical properties of epitaxial graphene films. The QHE is observed in gated epitaxial graphene films on SiC (0001), along with pronounced Shubnikov-de Haas (SdH) oscillations in magneto-transport.

The graphene films were grown on semi-insulating 4H-SiC substrates in an Epigress VP508 SiC hot-wall chemical vapor deposition (CVD) reactor. The off-cut angle of the substrate is nominally zero degrees. Prior to growth, substrates are subjected to a hydrogen etch at 1600 °C for 5 minutes, followed by cooling the samples to below 700 °C. After evacuating hydrogen from the system, the growth environment is pumped to an approximate pressure of $2\times10^{-7}$ mbar before temperature ramping at a rate of 10-20 °C/min and up to a specified growth temperature.

The growth conditions, film morphology, and electrical properties of the epitaxial graphene films differ markedly between films grown on the C-face and films grown on the Si-face. The formation of graphene on C-face is very rapid at the growth temperature of 1500-1650 °C in vacuum. At growth temperatures ≥ 1550 °C, several-micron large regions of smooth graphene films are obtained with tens of nanometers high ridges as domain boundaries.[16] Without gate stacks, the multi-layer graphene films on C-face are mostly p-type with a typical Hall mobility of 5000-6000 $cm^2/Vs$. On Si face, continuous few-layer graphene only starts to form at 1550 °C. The growth on Si-face is much slower, making it possible to form single layer graphene with a better controlled process.[17] The morphology of graphene films on the Si-face is quite homogenous compared to those films on the C-face. However, the typical Hall mobility of graphene on Si-face is ~ 1300-1600 $cm^2/Vs$. The particular graphene films shown in this Letter were grown at 1600 °C for 10 minutes in vacuum.

[a)] Author to whom correspondence should be addressed; electronic mail: yep@purdue.edu

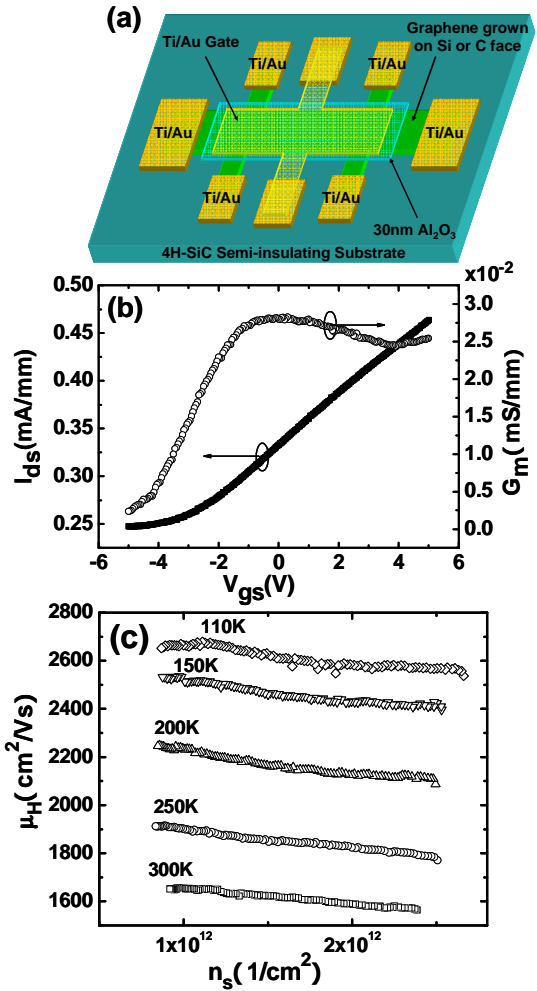

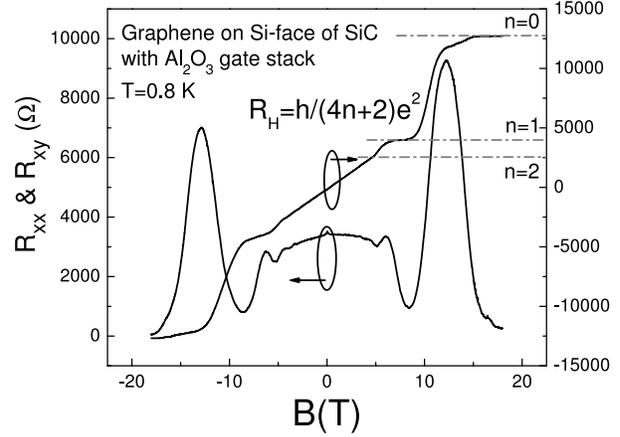

Figure 2 Hall resistance and magneto-resistance measured in the device in Figure 1(a) at T=0.8K and with floating gate bias. The horizontal dashed lines corresponding to $h/(4n+2)e^2$ values. The QHE of the electron gas in epitaxial graphene is shown one quantized plateau and two developing plateau in $R_{xy}$, with vanishing $R_{xx}$ in the corresponding magnetic field regime.

Figure 1(a) Schematic view of the graphene Hall-bar device structure on SiC (0001) with ALD $Al_2O_3$ as gate dielectric. (b) Drain current (left) and trans-conductance (right) versus gate bias at drain voltage of 50 mV for the fabricated graphene FET as shown in (a). (c) Hall mobility of the epitaxial graphene on SiC (0001) versus carrier density at various measurement temperatures.

The detailed device structure is shown in Fig. 1(a). The device isolation was achieved by $O_2$ plasma mesa dry etching. 1 nm of aluminum metal film was evaporated on the sample by electron-beam evaporation at ~ $10^{-6}$ Torr and fully oxidized in an oxygen rich ambient for 1 hour as a seeding layer for ALD growth. 30 nm $Al_2O_3$ as gate dielectric was deposited at growth temperature of 300 °C using an ASM F-120 reactor with tri-methyl aluminum and water vapor as the precursors. The metal contacts and gate electrodes were subsequently patterned and deposited, both using electron-beam evaporated Ti/Au. The active device area for magneto-transport has a width of 10 μm and a length of 22 μm. Four-point magneto-transport measurements are performed in a 1T Abbess Hall system (110K to 300K) or a variable temperature (0.4K to 70K) $^3$He cryostat in magnetic fields up to 18 T using standard low frequency lock-in technique. The external magnetic field (B) is applied normally to the graphene plane.

Figure 1(b) shows the dc $I_{ds}$-$V_{gs}$ and $G_m$-$V_{gs}$ characteristics with a gate bias from -5.0 to 5.0 V and $V_{ds}$=50 mV on a Hall-bar device with a partially-covered gate, as shown in Fig. 1(a). The left and right terminals of the Hall-bar serve as source and drain. The measured graphene field-effect transistor (FET) has a designed gate length ($L_g$) of 30 μm and a gate width ($L_w$) of 10 μm. The drain current can be modulated by ~46% with a few volts gate bias, similar to the previous work with $SiO_2$ as gate dielectric.[5] The device cannot be turned off due to the zero-bandgap of graphene films. The monotonic reduction in drain current with negative gate bias confirms that the carriers in graphene films on SiC (0001) are n-type. The slope of the drain current shows that the peak extrinsic transconductance ($G_m$) is ~ $2.8 \times 10^{-2}$ mS/mm, due to its extraordinarily large gate length and low drain bias. The channel mobility cannot be accurately measured by field-effect mobility $\mu=[(\Delta I_d/V_{ds})/(L_w/L_g)]/C_{ox}\Delta V_{gs}$ due to the following reasons. Here, $C_{ox}$ is determined by $\varepsilon_0\varepsilon_r A/d$, where $\varepsilon_0$ is the permittivity of free space, $\varepsilon_r$ is 8.0 for ALD $Al_2O_3$ without annealing, A is the unit area, and d is the gate oxide thickness. First, quantum capacitance plays a critical role for single-layer graphene[18] instead of multi-layer graphene[5]. Second, the real voltage across the modulated gate area is much smaller than $V_{ds}$ because of the partial gate design and the highly resistive area of graphene not covered by the gate. Third, the channel mobility could be under-estimated without correction of bulk and interface traps.[19] The most straightforward method to directly measure the mobility is the Hall measurement.

The Hall mobility and electron density of the graphene film are characterized at different gate biases and at different temperatures in Figure 1(c). The room temperature Hall mobility is ~ 1600 cm$^2$/Vs and decreases slightly with the increase of electron densities. There is also no significant Hall mobility degradation with gate stacks as compared to similar devices without gate dielectrics, indicating that the gate dielectric has reasonable quality without significant bulk traps and interface traps. The Hall mobility increases rapidly as temperature decreases and reaches ~2600 cm$^2$/Vs at 110K, and ~3600 cm$^2$/Vs at 4.2 K due to the suppression of electron-phonon scattering in epitaxial graphene films. So far the low-temperature mobility in epitaxial graphene is about three orders of magnitude lower than that in modulation-doped GaAs and also about one-two orders of magnitude lower than that in suspended exfoliated graphene. We ascribe it mainly due to the defects, impurities, and domain boundaries induced by the high-temperature sublimation process.

Figure 2 shows the magneto-resistance $R_{xx}$ and the Hall resistance $R_{xy}$ as a function of magnetic field $B$ from -18T to 18T at 0.8 K. From the Hall slope, the electron density is determined to be $1.04 \times 10^{12}$/cm$^2$ and Hall mobility of 3580 cm$^2$/Vs at 0.8 K. At high magnetic fields, $R_{xy}$ exhibits a plateau while $R_{xx}$ is vanishing, which is the fingerprint of the QHE and SdH oscillations. One well-defined plateau with value ($h/2e^2$) is observed at $|B|>15.5$T, while two higher-order plateaus are developing with values of ($h/6e^2$) and ($h/10e^2$), respectively. The pronounced SdH oscillations with at least four distinguishable peaks are also observed at the corresponding magnetic fields. The precision of the plateau is better than 1 part in $10^4$ within the instrumental uncertainty. It shows the QHE in epitaxial graphene is also applicable for metrology applications. The $R_{xy}$ quantization in this epitaxial graphene film is in accordance with [$h/(4n+2)e^2$], where $n$ is the Laudau level index, found in exfoliated graphene as a distinguishing feature of Dirac fermions. It is significantly different from conventional Fermi electrons with plateaus of ($h/ne^2$). The observed well-defined QHE reproduces the unique features observed in exfoliated single-layer graphene including a Berry phase of π. The observed QHE on this epitaxial graphene confirms that epitaxial graphene on SiC (0001) and exfoliated single-layer graphene are governed by the same relativistic physics with Dirac fermions as transport carries.[1-3] Due to the relatively thick graphene films on the C-face of SiC grown under similar conditions, no QHE is observed on C-face fabricated devices.

In conclusion, a high-$k$ gate stack on epitaxial graphene is realized by inserting a fully oxidized nanometer thin aluminum film as a seeding layer followed by an atomic-layer deposition process. The electrical properties of epitaxial graphene films are sustained after gate stack formation without significant degradation. At low temperatures, the QHE is observed in epitaxial graphene on SiC (0001), along with pronounced SdH oscillations. This quantum experiment confirms that epitaxial graphene on SiC (0001) shares the same relativistic physics as the exfoliated graphene. During the preparation and revision of this manuscript, we became aware of similar observations of the graphene-like QHE on both Si-face and C-face of SiC substrates. [20-22]

The authors would like to thank J.A. Cooper Jr., L.P. Rokhinson and A.T. Neal for valuable discussions, and G. Jones, T. Murphy and E. Palm at National High Magnetic Field Laboratory (NHMFL) for experimental assistance. Part of the work on graphene is supported by NRI (Nanoelectronics Research Initiative) through MIND (Midwest Institute of Nanoelectronics Discovery), DARPA and Intel Cooperation. NHMFL is supported by NSF Grant Nos. DMR-0084173 and ECS-0348289, the State of Florida, and DOE.

[1] A.K. Geim and K.S. Novoselov, *Nature Materials* **6**, 183 (2007).
[2] K.S. Novoselov, A.K. Geim, S.V. Morozov, D. Jiang, Y. Zhang, S.V. Dubonos, I.V. Grigorieva, and A.A. Firsov, *Science* **306**, 666 (2004).
[3] Y. Zhang, Y.W. Tan, H.L. Stormer, and P. Kim, *Nature* **438**, 201 (2005).
[4] C. Berger, Z. Song, X. Li, X. Wu, N. Brown, C. Naud, D. Mayou, T. Li, J. Hass, A.N. Marchenkov, E.H. Conrad, P.N. First, and W.A. de Heer, *Science* **312**, 1191 (2006).
[5] Y.Q. Wu, P.D. Ye, M.A. Capano, Y. Xuan, Y. Sui, M. Qi, J.A. Cooper, T. Shen, D. Pandey, G. Prakash, and R. Reifenberger, *Appl. Phys. Lett.* **92**, 092102 (2008).
[6] G. Gu, S. Niu, R.M. Feenstra, R.P. Devaty, W.J. Choyke, W.K. Chan, and M.G. Kane, *Appl. Phys. Lett.* **90**, 253507 (2007).
[7] J.S. Moon, D. Curtis, M. Hu, D. Wong, C. McGuire, P.M. Campbell, G. Jernigan, J.L. Tedesco, B. VanMil, R. Myers-Ward, C. Eddy, Jr., and D.K. Gaskill, *IEEE Electron Device Letters* **30**, 650 (2009).

[8] S.Y. Zhou, G.-H. Gweon, A.V. Fedorov, P.N. First, W.A. de Heer, D.-H. Lee, F. Guinea, A.H. Gastro Neto and A. Lanzara, Nature Materials 6, 770 (2007).
[9] J. Kedzierski, P.-L. Hsu, P.Healey, P.W. Wyatt, C.L. Keast, M. Sprinkle, C. Berger and W.A. de Heer, IEEE Trans. on Electron Devices 55, 2078 (2008).
[10] F. Varchon, R. Feng, J. Hass, X. Li, B.N. Nguyen, C. Naud, P. Mallet, J.-Y. Veuillen, C. Berger, E.H. Conrad, and L. Magaud, Phys. Rev. Lett. **99**, 126805 (2007).
[11] J. Hass, F. Vaechon, J.E. Millan-Otoya, M. Sprinkle, N. Sharma, W.A. de Heer, C. Berger, P.N. First, L. Magaud, and E.H. Conrad, Phys. Rev. Lett. **100**, 125504 (2008).
[12] Y. Xuan, Y. Wu, T. Shen, M. Qi, M.A. Capano, J.A. Cooper, and P.D. Ye, Appl. Phys. Lett. **92**, 013101 (2008).
[13] J.R. Williams, L. DiGarlo, and C.M. Marcus, *Science* **317**, 638 (2007).
[14] B. Lee, S.-Y. Park, H.-C. Kim, K.J. Cho, E.M. Vogel, M.J. Kim, R.M. Wallace, and J. Kim, *Appl. Phys. Lett.* **92**, 203102 (2008).
[15] S. Kim, J. Nah, I. Jo, D. Shahrjerdi, L. Colombo, Z. Yao, E. Tutuc, and S.K. Banerjee, *Appl. Phys. Lett.* **94**, 062107 (2008).
[16] L.B. Biedermann, M.L. Bolen, M.A. Capano, D. Zemlyanov, and R.G. Reifenberger, *Phys. Rev. B* **79**, 125411 (2009).
[17] J.A. Robinson, M. Wetherington, J.L. Tedesco, P.M. Campbell, X. Weng, J. Sttitt, M.A. Fanton, E. Frantz, D. Snyder, B.L. VanMil, G.G. Jernigan, R.L. Myers-Ward, C.R. Eddy, Jr., and D. K. Gaskill, *Nano Letters* **9**, 2873 (2009).
[18] Z. Chen and J. Appenzeller, in *IEDM Tech. Digest*, 509 (2008).
[19] Y. Xuan, Y.Q. Wu, H.C. Lin, T. Shen, and P.D. Ye, *IEEE Electron Dev. Lett.* 28, 935 (2007).
[20] J. Jobst, D. Waldmann, F. Speck, R. Hirner, D.K. Maude, T. Seyller, and H.B. Weber, arXiv:0908.1900v1.
[21] X. Wu, Y. Hu, M. Ruan, N.K. Madiomanana, J. Hankinson, M. Sprinkle, C. Berger, and W.A. de Heer, arXiv:0908.4112.
[22] A. Tzalenchuk, S. Lara-Avila, A. Kalaboukhov, S. Paolillo, M. Syvajarvi, R. Yakimova, O. Kazalova, T.J.B.M. Janssen, V. Fal'ko, and S. Kubatkin, arXiv:0909.1220.